%% file: paper.tex
\input 11layout
\input macro

\input epsf

\def\tld{\twoldots}

\centerline{\biggbf Finding Approximate Palindromes in Strings}

\bigskip\bigskip
\centerline{\it Alexandre H. L. Porto}
\centerline{\it Valmir C. Barbosa}

\bigskip
\centerline{Programa de Engenharia de Sistemas e Computa\c c\~ao, COPPE}
\centerline{Universidade Federal do Rio de Janeiro}
\centerline{Caixa Postal 68511}
\centerline{21945-970 Rio de Janeiro - RJ, Brazil}

\medskip
\centerline{\tt xandao@cos.ufrj.br, valmir@cos.ufrj.br}

\bigskip\bigskip
\centerline{\bf Abstract}

\medskip
\noindent
We introduce a novel definition of approximate palindromes in strings, and
provide an algorithm to find all maximal approximate palindromes in a string
with up to $k$ errors. Our definition is based on the usual edit operations of
approximate pattern matching, and the algorithm we give, for a string of size
$n$ on a fixed alphabet, runs in $O(k^2n)$ time. We also discuss two
implementation-related improvements to the algorithm, and demonstrate their
efficacy in practice by means of both experiments and an average-case analysis.


\vfill\eject
\bigbeginsection 1. Introduction

Let $S$ be a string of $n$ characters from a fixed alphabet $\Sigma$. For
$1\le i\le j\le n$, let $S[i]$ denote the $i$th character in $S$ and
$S[i\tld j]$ denote the substring of $S$ whose first and last characters are
$S[i]$ and $S[j]$, respectively. We let $S^R$ denote the string whose $i$th
character is $S[n-i+1]$, that is, $S$ and $S^R$ are essentially the same string
when read in opposing directions. We say that $S$ and $S^R$ are the
{\it reverse\/} of each other.

In this paper, we are concerned with {\it palindromes\/} occurring in $S$,
which are substrings $S[i\tld j]$ such that $S[i\tld j]=S[i\tld j]^R$. If
$S[i\tld j]$ is a palindrome, then it is an {\it even palindrome\/} if it
contains an even number of characters, otherwise it is an {\it odd palindrome}.
The {\it center\/} of this palindrome is $S[c]$, where
$c=i-1+\bigl\lceil(j-i+1)/2\bigr\rceil$. It follows that
$S[c+1\tld j]=S[i\tld c]^R$ for an even palindrome, while
$S[c+1\tld j]=S[i\tld c-1]^R$ for an odd palindrome. A palindrome $S[i\tld j]$
is an {\it initial palindrome\/} if $i=1$ and a {\it final palindrome\/} if
$j=n$. It is said to be a {\it maximal palindrome\/} if it is an initial
palindrome, a final palindrome, or if $S[i-1\tld j+1]$ is not a palindrome for
$1<i\le j<n$.

Several algorithms have appeared for the detection of palindromes in strings.
These include sequential algorithms for detecting all initial palindromes or
maximal palindromes centered at all positions [1, 2], as well as parallel
algorithms [3--6].

Palindromes appear in several domains, chiefly in computational molecular
biology [7, 8], where $\Sigma$ is for example the set of bases that link
together to form strands of nucleic acid. In this domain of application,
palindromes are often required to be {\it complementary}, in the following
sense. While for a palindrome centered at $c$ we have either $S[c-r+1]=S[c+r]$
or $S[c-r]=S[c+r]$, depending respectively on whether the palindrome is even or
odd and for $r$ representing distance from the center, if the palindrome is
complementary then the two characters are no longer equal but rather constitute
a complementary pair. In this case, we no longer have $S[i\tld j]=S[i\tld j]^R$,
but rather complementarity between the two strings.

Although we in the sequel deal exclusively with palindromes for which equality
is used to compare them with their reverses (everything carries over trivially
to the case of complementarity), we dwell a little longer on the application of
palindromes to computational molecular biology because in that area the
definition we have given for palindromes is ``too exact,'' being therefore of
little use. Palindromes that matter in that domain are ``approximate,'' in the
sense that the symmetry around a palindrome's center need not be perfect, but
may instead contain a certain number of mismatches in the form of gaps and
defects of various other natures [7, 8]. This paper is about finding
approximate palindromes in $S$.

The following is how the balance of this paper is organized. We start in
Section 2 by providing a precise definition of what is to be understood as
an approximate palindrome and a maximal approximate palindrome. This definition
is novel, as previous definitions appear to have been too restricted [7]. Then
in Section 3 we give an algorithm to find all maximal approximate palindromes
in a string allowing for $k$ errors. The algorithm runs in $O(k^2n)$ time.
Section 4 contains two improvements on the basic algorithm. These improvements
do not lead to an improved complexity, but do in practice make a difference,
as we demonstrate by means of some experimental results. Section 5 contains
an average-case analysis of the two improvements, which indicates that the
conclusions drawn in the previous section can be expected to hold on average.
Concluding remarks follow in Section 6.

\bigbeginsection 2. Approximate palindromes

Our definition of an approximate palindrome centered at position $c$ in string
$S$ is given for an integer $k\ge 0$ indicating the maximum number of errors to
be tolerated, and is based on the notion of {\it string editing}, that is,
the transformation of one string into another [7]. In order to define
approximate palindromes, we consider, for $1\le c\le n$, the editing of string
$S_\ell^c=S[1\tld u]^R$ to obtain string $S_r^c=S[c+1\tld n]$, where $u=c$ for
even palindromes or $u=c-1$ for odd palindromes (here, and henceforth,
$S[i\tld j]$ is to be understood as the empty string if $i>j$).

The string editing that we consider is the same that has been used for other
problems on strings, and employs the operations presented next. These operations
act on cursors $p$ and $q$, which are used to point to specific positions in
$S_\ell^c$ and $S_r^c$, respectively. Such cursors are such that $0\le p\le u$
and $0\le q\le n-c$, the value $0$ being used only to initialize the cursors as
still not pointing to characters in the string. Note that $p$ and $q$, when
nonzero, do not indicate positions in $S$, but in two of its substrings taken as
independent entities.

The operations we consider are the following, of which (ii)--(iv) are
called {\it edit operations}.

\medskip
\itemitem{(i)} {\it Matching\/}: If $S_\ell^c[p+1]=S_r^c[q+1]$, then
increment both $p$ and $q$.

\medskip
\itemitem{(ii)} {\it Substitution\/}: If $S_\ell^c[p+1]\neq S_r^c[q+1]$,
then increment both $p$ and $q$.

\medskip
\itemitem{(iii)} {\it Insertion\/}: Increment $q$.

\medskip
\itemitem{(iv)} {\it Deletion\/}: Increment $p$.

\medskip
While a matching can only be applied if it will make both cursors
point to equal characters, the edit operations characterize the
possible sources of mismatch between the two strings. What they do is to allow
for a character from $S_r^c$ to substitute for a character in $S_\ell^c$ (this
is a substitution), for a character from $S_r^c$ to be inserted into
$S_\ell^c$ as an additional character (an insertion), or for a character to be
deleted from $S_\ell^c$ (a deletion). When grouped into a sequence, what
matching and edit operations can be regarded as doing is providing a script
(the {\it edit script\/}) for a prefix of $S_\ell^c$ to be turned into a
replica of a prefix of $S_r^c$ (note that, if an unlimited number of edit
operations is allowed, then such an edit script is guaranteed to exist).
The edit script to convert one prefix into the other having the smallest
number of edit operations is said to be {\it optimal\/} for the two prefixes,
and its number of edit operations is called the {\it edit distance\/} between
them [9, 10].

For $0\le p^*\le u$ and $0\le q^*\le n-c$, we say that $S[u-p^*+1\tld c+q^*]$
is an {\it approximate palindrome\/} in $S$ centered at $c$ if, of
all the edit scripts that can be used to turn $S_\ell^c[1\tld p^*]$ into
$S_r^c[1\tld q^*]$, the one that is optimal comprises no more than $k$ edit
operations (in other words, the edit distance between $S_\ell^c[1\tld p^*]$ and
$S_r^c[1\tld q^*]$ is at most $k$). In this case, $p^*$ and $q^*$ are the values
of $p$ and $q$, respectively, after any of those edit scripts is played out.
The palindrome is even or odd according to how $S_\ell^c$ is originally set.
Its {\it size\/} is either $p^*+q^*$ or $p^*+q^*+1$, respectively if it is even
or odd. This definition is more general than the single other definition that
appears to have been given for approximate palindromes [7], which only allows
for matchings and substitutions.

It is curious to observe that, unlike exact palindromes, the size of an even
approximate palindrome does not have to be even, nor does the size of an odd
approximate palindrome have to be odd. These would hold, however, for the exact
palindrome that would be obtained if the transformation of one prefix into the
other were indeed performed.

We provide in Figure 1 an illustration of this concept of an approximate
palindrome. The figure contains four approximate palindromes in the string
${\it bbaabac}$ for $k=3$, two even in part (a) and two odd in part (b).
The palindromes are depicted in a way that evidences their two parts and also
their centers, in the odd case. Blank cells appearing in the left part
correspond to insertions, those on the right to deletions (for ease of
representation, in the figure we let deletions from the left part be
represented as insertions into the right part). Substitutions are indicated
by shaded cells placed symmetrically in the two parts.

Having defined approximate palindromes for $c$ and $k$ fixed, the notion that
remains to be introduced before we move on to discuss their detection is that
of maximality. Note, first, that the simple definition of a maximal palindrome
in the exact case does not carry over simply to the approximate case. In the
case of exact palindromes, maximality is related to the inability to extend a
palindrome into another substring of $S$ that is also a palindrome. In the
approximate case, however, the potential existence of several acceptable edit
scripts makes it inappropriate to adopt such a straightforward definition.

While there does appear to exist more than one possibility for defining the
maximality of approximate palindromes, what we do in this paper is to say
that an approximate palindrome is {\it maximal\/} if no other
approximate palindrome for the same $c$ and $k$ exists having strictly greater
size or the same size but strictly fewer errors (edit distance between its two
parts). Unlike the case of exact palindromes, this definition clearly does not
guarantee the uniqueness of an approximate palindrome that is maximal.

\bigbeginsection 3. An algorithm

In this section, and for $k$ fixed, we introduce an algorithm for detecting
maximal approximate palindromes, one even and one odd for each $c$ such that
$1\le c\le n$.

Our algorithm is based on an acyclic directed graph $D$, whose definition
relies on two generic strings $X$ and $Y$ on the same alphabet. We let
$x=\vert X\vert$ and $y=\vert Y\vert$. Graph $D$ has $(x+1)(y+1)$ nodes, one
for each $(i,j)$ pair such that $0\le i\le x$ and $0\le j\le y$. A directed
edge exists in $D$ from node $(i,j)$ to node $(i',j')$ if either $i'=i$
and $j'=j+1$, or $i'=i+1$ and $j'=j$, or yet $i'=i+1$ and $j'=j+1$. If we
position the nodes of $D$ on the vertices of a two-dimensional grid having $x+1$
rows and $y+1$ columns so that a node's first coordinate grows from top to
bottom and the second from left to right, then clearly directed edges exist
between nearest neighbors in the same row (a {\it horizontal edge\/}),
the same column (a {\it vertical edge\/}), and the same diagonal (a
{\it diagonal edge\/}). Because $j'-i'=j-i$ when a diagonal edge exists from
$(i,j)$ to $(i',j')$, we use such differences to label the various diagonals on
the grid. Diagonal labels are then in the range of $-x$ through $y$.

Now consider a directed path in $D$ leading from node $(i,j)$ to node $(i',j')$.
By definition of the directed edges, clearly $i'\ge i$ and $j'\ge j$. The
importance of graph $D$ in our present context is that this directed path, if
it contains at least one edge, can be interpreted as an edit script for string
$X[i+1\tld i']$ to be transformed into string $Y[j+1\tld j']$. Along the script,
the cursors $p$ and $q$ of operations (i)--(iv) are used on $X$ and $Y$,
respectively, such that $i\le p\le i'$ and $j\le q\le j'$. On such a path, a
diagonal edge corresponds to either a matching or a substitution, a horizontal
edge to an insertion, and a vertical edge to a deletion. The number of edges on
the path that do not correspond to matchings is the number of edit operations in
the script. The optimal edit script for strings $X[i+1\tld i']$ and
$Y[j+1\tld j']$ is represented in $D$ by a directed path from $(i,j)$ to
$(i',j')$ whose number of edges corresponding to edit operations is minimum
among all directed paths between the two nodes. Such a path is said to be
{\it shortest\/} among all those paths according to the metric that assigns,
say, length $0$ to edges corresponding to matchings and length $1$ to all other
edges.

We give an illustration in Figure 2, where graph $D$ is shown for
$X={\it bb}$ and $Y={\it aabac}$. The directed path shown in solid lines
contains two insertions, followed by one matching and one substitution,
and is a shortest path between its two end vertices. It corresponds,
therefore, to an optimal edit script to transform $X[1\tld 2]$ into
$Y[1\tld 4]$. All other edges are shown as dotted lines, directions omitted
for clarity.

For $0\le e\le k$, let a directed path in $D$ be called an {\it $e$-path\/} if
it contains $e$ edges related to edit operations. One crucial problem to be
solved on $D$ is the problem of determining, for each diagonal and each $e$, the
$e$-path, if one exists, that starts in row $0$, ends on that diagonal at the
farthest possible node (greatest row number), and is in addition shortest among
all paths that start and end at the same nodes. This problem can be solved by
the following dynamic-programming approach.

\medskip
\itemitem{1.} For $d=0,\ldots,y$, find the largest common prefix of strings
$X$ and $Y[d+1\tld y]$. Such prefixes will correspond to the $0$-paths that
start in row $0$ and end at the farthest possible nodes. Each of them will
be entirely confined to a diagonal $d$ and, like all $0$-paths, will be
shortest among all paths joining its end nodes.

\medskip
\itemitem{2.} For $e=1,\ldots,k$, and for $d=-\min\{e,x\},\ldots,y$, do:

\smallskip
\itemitemitem{2.1.} Consider the $(e-1)$-paths, if any,  determined in the
previous step on those of diagonals $d-1$, $d$, and $d+1$ that exist. Each of
these paths corresponds to an optimal edit script for transforming a prefix of
$X$ into a substring of $Y$. If possible, extend these scripts, respectively
by adding an insertion, a substitution, and a deletion, thereby creating
$e$-paths that end on diagonal $d$.

\smallskip
\itemitemitem{2.2.} Of the $e$-paths created in step 2.1, if any, pick the
one that ends farthest down diagonal $d$ and extend it further by computing the
largest common prefix of what remains of $X$ and what remains of $Y$. The result
will be an $e$-path that starts in row $0$ and ends on diagonal $d$ at the
farthest possible node, being in addition shortest among all paths starting and
ending at the same nodes.

\medskip
The lower bound on $d$ in step 2 reflects the fact that diagonal $-e$ can only
be reached from row $0$ by an $e'$-path, where $1\le e\le e'\le k$. What the
entire procedure computes is a set of directed paths departing from row $0$ at
several columns. Each of these directed paths is an $e$-path, for some $e$ such
that $0\le e\le k$, that ends as far down in the graph as possible, and is also
shortest among all paths that start and end at the same nodes. So an $e$-path
computed by the algorithm departing from node $(0,d)$ for $0\le d\le y$
represents an optimal edit script for turning a prefix of $X$ into a prefix of
$Y[d+1\tld y]$ with $e$ edit operations.

The basic procedure comprising steps 1 and 2 was introduced to solve the problem
of approximate pattern matching [7, 11--13], which requires all approximate
occurrences of $X$ in $Y$ having edit distance from $X$ of at most $k\le y$ to
be determined [14]. The solution works by selecting, after the
execution of steps 1 and 2, the $e$-paths that end on row $x$ for $0\le e\le k$.
Several other solutions to this problem exist [7, 15--20].

Note that the largest common prefixes asked for in steps 1 and 2.2 can be
obtained easily if we have a means of computing the largest common prefix of any
two suffixes of $X\$_1Y$, where $\$_1$ is any character that does not occur in
$X$ or $Y$, and $X\$_1Y$ is the string obtained by appending $\$_1$ to $X$, then
$Y$ to $X\$_1$. Such a means, of course, is provided by the well-known suffix
tree for string $X\$_1Y\$_2$, where $\$_2$ is any character not occurring in
$X\$_1Y$, ultimately needed to ensure that the tree does indeed exist. After
the suffix tree is built and preprocessed, which can be achieved in $O(x+y)$
time, any of those largest common prefixes can be found in constant time [7].
Fox approximate pattern matching, $x\le y$, so $O(y)$ is the time it takes to
handle the suffix tree initially. After that, the complexity of steps 1 and 2
is dominated by step 2, which comprises $O(ky)$ repetitions of 2.1 and 2.2,
each requiring $O(1)$ time per repetition. The overall time is then $O(ky)$.

The same basic procedure can also be used to solve another problem involving
strings $X$ and $Y$, known as the $k$-differences problem. Assuming
$k\le\max\{x,y\}$ to avoid trivial cases, this problem asks for the edit script
to transform $X$ into $Y$ with the fewest possible edit operations, but no more
than $k$ [13, 21]. Clearly, no solution exists if $\vert x-y\vert>k$. A solution
may exist otherwise, and will correspond to the $e$-path from $(0,0)$ to $(x,y)$
for which $e$ is minimum (that is, a shortest path between the two nodes), if
one exists with $e\le k$. Adapting steps 1 and 2 to find such a path is a simple
matter, as follows. In step 1, let $d=0$ only. In step 2, let the range for $d$
be from $-\min\{e,x\}$ through $\min\{e,y\}$, again reflecting the fact that it
takes at least $e$ errors to reach diagonal $-e$ or $e$ from $(0,0)$. Finally,
abort the iterations whenever node $(x,y)$ is reached.

This solution to the $k$-differences problem requires a number of matching
extensions (steps 1 and 2.2) given by $1+\sum_{e=1}^kO(e)=O(k^2)$, each one
requiring $O(1)$ time after the initial construction and preprocessing of the
suffix tree for $X\$_1Y\$_2$ in $O(x+y)$ time. The overall time is then
$O(k^2+x+y)$.

This algorithm for the $k$-differences problem can be used directly to solve
our problem of determining all maximal approximate palindromes in $S$, because
what is required of an approximate palindrome is precisely that one of its parts
be transformable into the other by means of an optimal edit script comprising
no more than $k$ edit operations. For $1\le c\le n$, we simply let $X=S_\ell^c$
and $Y=S_r^c$ (then $x+y=n$ or $x+y=n-1$, respectively for even and odd
palindromes). Whenever a new path is determined in step 2.2, we check if it is
better than the ones found previously in terms of approximate-palindrome
maximality; it will be better if it leads to a node whose coordinates add to a
larger integer than the current best path (checking for the same integer but
fewer errors is needless, as the algorithm generates shortest $e$-paths in
nondecreasing order of $e$).

Apart from the time needed to establish the suffix tree for
$S_\ell^c\$_1S_r^c\$_2$, this algorithm requires $O(k^2)$ time to determine a
maximal approximate palindrome in $S$ for fixed $c$. Determining all even and
odd approximate palindromes in $S$ then requires $O(k^2n)$ time beyond what is
needed to establish the suffix trees. If such a tree had indeed to be
constructed and preprocessed for each $c$, then an additional $O(n^2)$ time
would be required. However, note that every suffix of $S_\ell^c$ is also a
suffix of $S^R$, and that every suffix of $S_r^c$ is also a suffix of $S$. So
all that is required by steps 1 and 2.2, regardless of the value of $c$, is that
a preprocessed suffix tree be available for $S^R\$_1S\$_2$. This tree needs to
be established only once, which can be done in $O(n)$ time, and therefore the
overall complexity of determining all even and odd maximal approximate
palindromes in $S$ remains $O(k^2n)$.

Assessing the space required by the algorithm depends on whether edit scripts
are also needed or simply the palindromes with the corresponding edit distances.
In the former case, the space required for each of the $O(k)$ diagonals is
$O(k)$; it is constant in the latter case. This, combined with the $O(n)$ space
needed for preprocessing, yields a space requirement of $O(k^2+n)$ or $O(k+n)$,
respectively.

\bigbeginsection 4. Practical improvements

Of the $2n$ maximal approximate palindromes determined by the algorithm of
Section 3 for $c=1,\ldots,n$, three are trivial and can be skipped in a
practical implementation. These are the odd palindrome for $c=1$, and the
even and odd palindromes for $c=n$. For these palindromes, the optimal edit
scripts contain $k$ insertions for $c=1$ and $k$ deletions for $c=n$.
Henceforth in the paper, we then assume that the algorithm is run for
$c=1,\ldots,n-1$ in the even case, $c=2,\ldots,n-1$ in the odd case.

In addition to this simplification, the algorithm introduced in Section 3 for
the computation of all maximal approximate palindromes in $S$ can be improved
by selecting the range for $d$ in step 2 more carefully. We discuss two such
improvements in this section. Although they do not lead to a better execution
time in the asymptotic, worst-case sense, they do bring about a reduction in
execution times in practice, as we demonstrate.

The first improvement consists of skipping the diagonals on which $e$-paths,
for suitable $e$, have been determined that end on row $x$ or column $y$.
The reason why this is safe to do is that no further path can be found ending
on those diagonals farther down from row $x$ or column $y$. The way to
efficiently handle this improved selection of diagonals in step 2 is to keep
all diagonals that are going to be processed in a doubly-linked list. This
allows new diagonals to be added at the list's extremes in constant time, and
diagonals that will no longer be processed can be deleted equally efficiently
as soon as it is detected that the corresponding $e$-paths have reached the
farthest row or column.

The second improvement that we describe is itself an improvement over the
first one. The rationale is that, if a diagonal $d$ is dropped from further
consideration because the $e$-path that ends on it farthest down the grid
touches, say, row $x$, then all other diagonals $d'$ such that $d'<d$ may be
dropped as well. Similarly if that $e$-path on $d$ touches column $y$, in which
case not only diagonal $d$ but also all diagonals $d'$ such that $d'>d$ may be
dropped from further processing. What results from this improvement is that the
algorithm's operation is always confined to within a strip of contiguous
diagonals. This strip is limited on the left by either the leftmost diagonal on
which the farthest-reaching $e$-path does not touch row $x$ or diagonal
$-\min\{e,x\}$; on the right, it is limited by either the rightmost diagonal on
which the farthest-reaching $e$-path does not touch column $y$ or diagonal
$\min\{e,y\}$. As with the first improvement, it is a simple matter to implement
the second one efficiently by using the same doubly-linked list.

We show in Figures 3 and 4 the gain that the first improvement elicits for a
number of strings, in Figures 5 and 6 the gain due to the second improvement,
and in Figures 7 and 8 the gain caused by the second improvement over the first.
For Figures 3 through 6, if $t_o$ is the the number of iterations performed by
the original algorithm and $t_i$ the number of iterations performed by the
improved algorithm, then gain is defined as $(t_o-t_i)/t_o$. An iteration is
either the initial execution of step 1 or each combined execution of steps
2.1 and 2.2. Because every iteration can be carried out in constant time, our
assessment of gain in terms of numbers of iterations as opposed to elapsed time
provides a platform-independent evaluation of the improved algorithms. Gain is
defined similarly for Figures 7 and 8.

The strings we have used are the ones given next. Of these, some are
{\it periodic}, meaning that there exists a string $P$ of size $s\le n$ such
that the periodic string is a prefix of the string formed by concatenating
$\lceil n/s\rceil$ copies of $P$. The {\it period\/} of the periodic string
is the value of $s$.

\medskip
\itemitem{$\bullet$} ${\it dna}$: A DNA sequence with $n$ bases.

\medskip
\itemitem{$\bullet$} ${\it dnap1}$: A periodic DNA sequence with $n$
bases and period $\lfloor 0.05n\rfloor$.

\medskip
\itemitem{$\bullet$} ${\it dnap2}$: A periodic DNA sequence with $n$
bases and period $\lfloor 0.25n\rfloor$.

\medskip
\itemitem{$\bullet$} ${\it dnap3}$: A periodic DNA sequence with $n$
bases and period $\lfloor 0.5n\rfloor$.

\medskip
\itemitem{$\bullet$} ${\it txt}$: A sequence of $n$ ASCII characters.

\medskip
\itemitem{$\bullet$} ${\it txtp1}$: A periodic sequence of $n$ ASCII
characters with period $\lfloor 0.05n\rfloor$.

\medskip
\itemitem{$\bullet$} ${\it txtp2}$: A periodic sequence of $n$ ASCII
characters with period $\lfloor 0.25n\rfloor$.

\medskip
\itemitem{$\bullet$} ${\it txtp3}$: A periodic sequence of $n$ ASCII
characters with period $\lfloor 0.5n\rfloor$.

\medskip
\itemitem{$\bullet$} ${\it cnst}$: An $n$-fold repetition of the same
character.

\medskip
\itemitem{$\bullet$} ${\it diff}$: A string comprising $n$ different
characters. This string does not fit the fixed-alphabet assumption we have made
from the start, so the complexity figures given in Section 3 do not apply to
executions of the algorithm on it.

\medskip
Figures 3, 5, and 7 are given for $n=50$, while Figures 4, 6, and 8 are for
$n=2500$. In each figure, the values of $k$, the maximum number of errors to be
allowed in the approximate palindromes, are in
$\bigl\{\lceil 0.01n\rceil,
\lceil 0.05n\rceil, \lceil 0.1n\rceil, \lceil 0.2n\rceil,
\lceil 0.4n\rceil, \lceil 0.8n\rceil\bigr\}$. Note, in Figures 3, 5, and 7,
that gains are identical for ${\it dnap1}$ and ${\it txtp1}$, owing to the
fact that, for $n=50$, these two strings are essentially the same, being
periodic with period $2$, therefore having only $2$ different characters
throughout.

With the single exception of Figure 7, we see in all cases that gains are
largest for ${\it cnst}$, which can be easily accounted for by the fact that,
for such a string, every diagonal is dropped from further consideration by any
of the two improvements right after having been processed for the first time.
The exception of Figure 7 can be explained by the fact that this figure gives
gains of one improvement over the other, and together with Figure 8 gives
a measure for how efficiently the second improvement creates the diagonal
strips. What Figure 7 indicates is that such an efficiency is higher for
${\it dnap1}$ and ${\it txtp1}$ than it is for ${\it cnst}$ if $n=50$ and
$k\ge 3$.

By contrast, gains are always smallest for ${\it diff}$, because matchings
never occur in the edit scripts and therefore
more iterations are needed before paths reach the grid's borders. For
${\it diff}$ strings, it also happens that the second improvement provides no
gain over the first.

Gains tend to be larger for sequences on smaller alphabets, because in
these cases there tend to be more matchings. This is what happens for DNA
sequences, which have an alphabet of size $4$, therefore smaller than the
alphabets of nearly all the other strings under consideration. Gains also
tend to be larger as $k$ gets larger, which is probably related to the fact
that the overall number of iterations also increases with $k$. To finalize,
we note that gains tend to decrease as $n$ is increased, which can be
seen by comparing Figures 3, 5, and 7 to Figures 4, 6, and 8, respectively.
This is due to the fact that, as $n$ gets larger, so does the number of
matchings needed for paths to reach the grid's bottommost or
rightmost border.

\bigbeginsection 5. Expected gains

As we know from the results in Section 4, one of the key factors affecting the
performance of practical implementations of our algorithm is the size of the
alphabet $\Sigma$, which henceforth we let be such that
$\sigma=\vert\Sigma\vert$. We discuss, in this section, a stochastic model that
can be used to assess, to a limited extent, the expected gain provided by the
two improvements of Section 4 for a given value of $\sigma$. The model has great
richness of detail [22], and may require considerable computational effort to be
solved even for strings comprising as small a number of characters as a few
tens. It is therefore not practical, but we present an outline of it nonetheless
because, already for a modest range of parameters, it is capable of conveying
useful information.

For fixed $c$, the model views an execution of the algorithm (as introduced
originally or in one of its two improved variants), as a discrete-time,
discrete-state stochastic process. Each time step in this stochastic process
corresponds to one of the iterations on $e$ in step 2 of the algorithm. Each
state is a $(2k+1)$-tuple with one entry for each of the possible diagonals
from $-k$ through $k$. An entry contains the number of the row to which the
corresponding diagonal has been stretched so far by the algorithm. The initial
state has $0$ in all entries.

This stochastic process never returns to a previously visited state, and can
as such be represented by an acyclic directed graph whose nodes stand for
states and whose directed edges represent the possible transitions among states.
In this graph, every state is on some directed path from the initial state.
Associated with an edge $(P,Q)$ are two quantities, namely the probability
$p(P,Q)$ of moving from state $P$ to state $Q$, and the number of diagonals
to be processed when that transition is undertaken. This number of diagonals
is, for each value of $e$, the number of iterations used in Section 4 to
evaluate the two improvements as a platform-independent measure of time. It
depends on whether the original algorithm or one of its two improved variants
is in use, and is denoted by $t(P,Q)$.

The crucial issue in setting up this stochastic model is of course the
determination of $p(P,Q)$ for all edges $(P,Q)$. As it turns out, such
probabilities depend on how state $P$ is reached from the initial state,
and therefore it is best to assess them dynamically as the graph is processed
during the computation of one of the stochastic process' characteristics.
The characteristic that interests us in this section is the expected (average)
number of iterations, denoted by $\bar t$, for the algorithm to be completed on
a string, $n$ and $\sigma$ being fixed in addition to $c$.

The following recursive procedure is a variation of straightforward depth-first
traversal, and can be used to compute $\bar t$. It is started by executing step
1 on the initial state; upon termination, its output is assigned to $\bar t$.

\medskip
\itemitem{1.} Let $P$ be the current state. If no edges outgo from $P$ in the
graph, then return $0$. Otherwise, for $z>0$, let $Q_1\ldots,Q_z$ be the states
to which a directed edge outgoes from $P$. Do:

\smallskip
\itemitemitem{1.1.} For $i=1,\ldots,z$, recursively execute step 1 on state
$Q_i$, and let $t_i$ be its output.

\smallskip
\itemitemitem{1.2.} For $i=1,\ldots,z$, assess the transition probability
$p(P,Q_i)$ as a function of the directed path through which $P$ was reached.

\smallskip
\itemitemitem{1.3.} Return $\sum_{i=1}^z\bigl(t(P,Q_i)+t_i\bigr)p(P,Q_i)$.

\medskip
What remains to be presented before we discuss some results is, naturally, how
to perform step 1.2 of this procedure. First we set up two matrices, called
$M(P)$ and $M(Q_i)$, representing respectively the relationships (equality,
inequality, or none) that must exist between characters of two generic strings
$S_\ell^c$ and $S_r^c$ for state $P$ to be reached from the initial state on the
specific path that is being considered, and for $Q_i$ to be reached on that same
path elongated by the edge $(P,Q_i)$. If $a$ stands for the number of different
strings $S$ satisfying the constraints of $M(P)$ and $b$ the number of those
that satisfy the constraints of $M(Q_i)$, then $p(P,Q_i)=b/a$.

The computation of $a$ and $b$ can be reduced to a complex combinatorial
problem, of which we give an outline next, and constitutes the computationally
hardest part of the entire procedure to compute $\bar t$. Suppose we wish to
compute $a$ from $M(P)$. The way to proceed is to start by setting up an
undirected graph, call it $G$, whose nodes represent groups of positions in the
two parts of $S$ that by $M(P)$ must contain the same character. Two nodes are
connected by an edge if the corresponding groups of positions must, again by
$M(P)$, contain characters that differ from one group to the other. The value
of $a$ is then the number of distinct ways in which we can assign characters
from $\Sigma$ to the nodes of $G$ in such a way that nodes that are
connected by an edge receive different characters. This is a graph coloring
problem, that is, a problem related to assigning objects (colors) to the nodes
of a graph so that nodes that are connected by an edge receive different
objects. The problem is to find out the number of ways in which $G$'s
nodes can be colored by a total of $\sigma$ distinct colors. This number is
given by the so-called chromatic polynomial of $G$ evaluated at $\sigma$ [23].

Finally, we present some numerical results. These are illustrated in Figures 9
through 12. Of these, Figures 9 and 11 are for $k=1$ ($n$ and $\sigma$ vary),
while Figures 10 and 12 are for $n=10$ ($k$ and $\sigma$ vary). What the
figures depict are the gains, averaged over $c$ for both even and odd
palindromes, that correspond to the expected times $\bar t$ assessed for the
original algorithm and its two variants. Figures 9 and 10 give gains for the
first improvement, and Figures 11 and 12 for the second. These figures tend to
support the conclusions we drew from specific examples in Section 4. These are
that gains are expected to be larger for larger $k$ and smaller $\sigma$, and
smaller for larger $n$.

\bigbeginsection 6. Concluding remarks

We have in this paper introduced a novel definition of approximate palindromes
in strings, and have given an algorithm for finding all approximate palindromes
in a string of $n$ characters to within at most $k$ errors. For a fixed
alphabet, the algorithm runs in time $O(k^2n)$. We have also indicated how to
perform implementation-related improvements in the algorithm, and demonstrated,
over a variety of strings and also based on an average-case analysis, that such
improvements do indeed often lead to reduced running times in practice.

\beginsection  Acknowledgments

The authors have received partial support from the Brazilian agencies CNPq and
CAPES, the PRONEX initiative of Brazil's MCT under contract 41.96.0857.00, and
a FAPERJ BBP grant.

\bigbeginsection References

{\frenchspacing

\medskip
\item{1.} D. E. Knuth, J. H. Morris, and V. R. Pratt,
``Fast pattern matching in strings,''
{\it SIAM J. on Computing\/} {\bf 6} (1977), 322--350.

\medskip
\item{2.} G. Manacher,
``A new linear-time on-line algorithm for finding the smallest initial
palindrome of a string,''
{\it J. of the ACM\/} {\bf 22} (1975), 346--351.

\medskip
\item{3.} A. Apostolico, D. Breslauer, and Z. Galil,
``Optimal parallel algorithms for periods, palindromes and squares,''
{\it Proc. of the Int. Colloq. on Automata, Languages, and Programming},
296--307, 1992.

\medskip
\item{4.} A. Apostolico, D. Breslauer, and Z. Galil,
``Parallel detection of all palindromes in a string,''
{\it Theoretical Computer Science\/} {\bf 141} (1995), 163--173.

\medskip
\item{5.} D. Breslauer and Z. Galil,
``Finding all periods and initial palindromes of a string in parallel,''
{\it Algorithmica\/} {\bf 14} (1995), 355--366.

\medskip
\item{6.} Z. Galil,
``Optimal parallel algorithms for string matching,''
{\it Information and Control\/} {\bf 67} (1985), 144--157.

\medskip
\item{7.} D. Gusfield,
{\it Algorithms on Strings, Trees, and Sequences: Computer Science and
Computational Biology},
Cambridge University Press, New York, NY, 1997.

\medskip
\item{8.} J. Jurka,
``Origin and evolution of Alu repetitive elements,''
in R. J. Maraia (Ed.),
{\it The Impact of Short Interspersed Elements (SINEs) on the Host Genome},
R. G. Landes, New York, NY, 25--41, 1995.

\medskip
\item{9.} V. I. Levenstein,
``Binary codes capable of correcting insertions and reversals,''
{\it Soviet Physics Doklady\/} {\bf 10} (1966), 707--710.

\medskip
\item{10.} D. Sankoff and J. Kruskal (Eds.),
{\it Time Warps, String Edits, and Macromolecules: The Theory and Practice of
Sequence Comparison},
Addison-Wesley, Reading, MA, 1983.

\medskip
\item{11.} G. M. Landau and U. Vishkin,
``Introducing efficient parallelism into approximate string matching and a new
serial algorithm,''
{\it Proc. of the Annual ACM Symp. on Theory of Computing}, 220--230, 1986.

\medskip
\item{12.} G. M. Landau and U. Vishkin,
``Fast parallel and serial approximate string matching,''
{\it J. of Algorithms\/} {\bf 10} (1989), 157--169.

\medskip
\item{13.} E. W. Myers,
``An $O(nd)$ difference algorithm and its variations,''
{\it Algorithmica\/} {\bf 1} (1986), 251--266.

\medskip
\item{14.} E. Ukkonen,
``Algorithms for approximate string matching,''
{\it Information and Control\/} {\bf 64} (1985), 100--118.

\medskip
\item{15.} R. Baeza-Yates and G. Navarro,
``Faster approximate string matching,''
{\it Algorithmica\/} {\bf 23} (1999), 127--158.

\medskip
\item{16.} W. I. Chang and J. Lampe,
``Theoretical and empirical comparisons of approximate string matching
algorithms,''
{\it Proc. of the Symp. on Combinatorial Pattern Matching},
175--184, 1992.

\medskip
\item{17.} R. Cole and R. Hariharan,
``Approximate string matching: a simpler faster algorithm,''
{\it Proc. of the Annual ACM-SIAM Symp. on Discrete Algorithms},
463--472, 1998.

\medskip
\item{18.} G. A. Stephen,
{\it String Searching Algorithms},
World Scientific, Singapore, 1994.

\medskip
\item{19.} E. Ukkonen,
``Finding approximate patterns in strings,''
{\it J. of Algorithms\/} {\bf 6} (1985), 132--137.

\medskip
\item{20.} S. Wu and U. Manber,
``Fast text searching allowing errors,''
{\it Comm. of the ACM\/} {\bf 35} (1992), 83--91.

\medskip
\item{21.} G. M. Landau and U. Vishkin,
``Efficient string matching with $k$ mismatches,''
{\it Theoretical Computer Science\/} {\bf 43} (1986), 239--249.

\medskip
\item{22.} A. H. L. Porto,
{\it Detecting Approximate Palindromes in Strings},
M.Sc. Thesis, Federal University of Rio de Janeiro, Rio de Janeiro, Brazil,
1999 (in Portuguese).

\medskip
\item{23.} J. A. Bondy and U. S. R. Murty,
{\it Graph Theory with Applications},
North-Holland, New York, NY, 1976.

}

\beginsection Authors' biographical data

\vskip-\smallskipamount\medskip\noindent
{\bf Alexandre H. L. Porto} is a doctoral student at the Systems Engineering
and Computer Science Program of the Federal University of Rio de Janeiro. He
is interested in sequential and parallel algorithms for problems in
computational biology.

\medskip
\noindent
{\bf Valmir C. Barbosa} is professor at the Systems Engineering and Computer
Science Program of the Federal University of Rio de Janeiro, and is interested
in the various aspects of distributed and parallel computing, as well as of
the so-called complex systems, like neural networks and related models. He
received his Ph.D. from the University of California, Los Angeles, in 1986,
and has held visiting positions at the IBM Rio Scientific Center in Brazil,
the International Computer Science Institute in Berkeley, and the Computer
Science Division of the University of California, Berkeley. He has authored
the books {\it Massively Parallel Models of Computation\/} (Ellis Horwood,
Chichester, UK, 1993), {\it An Introduction to Distributed Algorithms\/} (The
MIT Press, Cambridge, MA, 1996), and {\it An Atlas of Edge-Reversal Dynamics\/}
(Chapman \& Hall/CRC Press, London, UK, 2000).

\vfill\eject

\topinsert
\centerline{\epsfbox{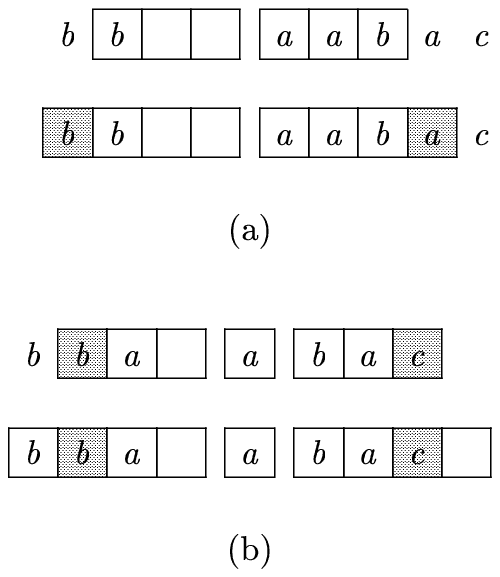}}
\bigskip
\centerline{{\bf Figure 1.} Even (a) and odd (b) approximate palindromes for
$k=3$ in the string ${\it bbaabac}$}
\bigskip\bigskip\bigskip
\endinsert

\topinsert
\centerline{\epsfbox{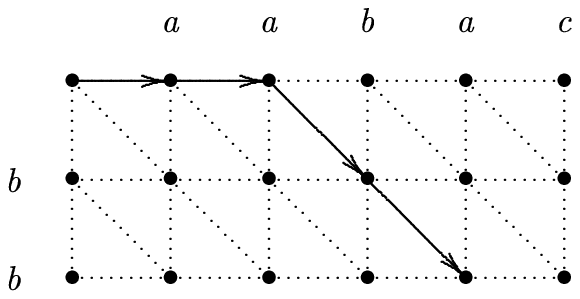}}
\bigskip
\centerline{{\bf Figure 2.} An edit script as a directed path in $D$}
\bigskip\bigskip\bigskip
\endinsert

\topinsert
\centerline{\epsfbox{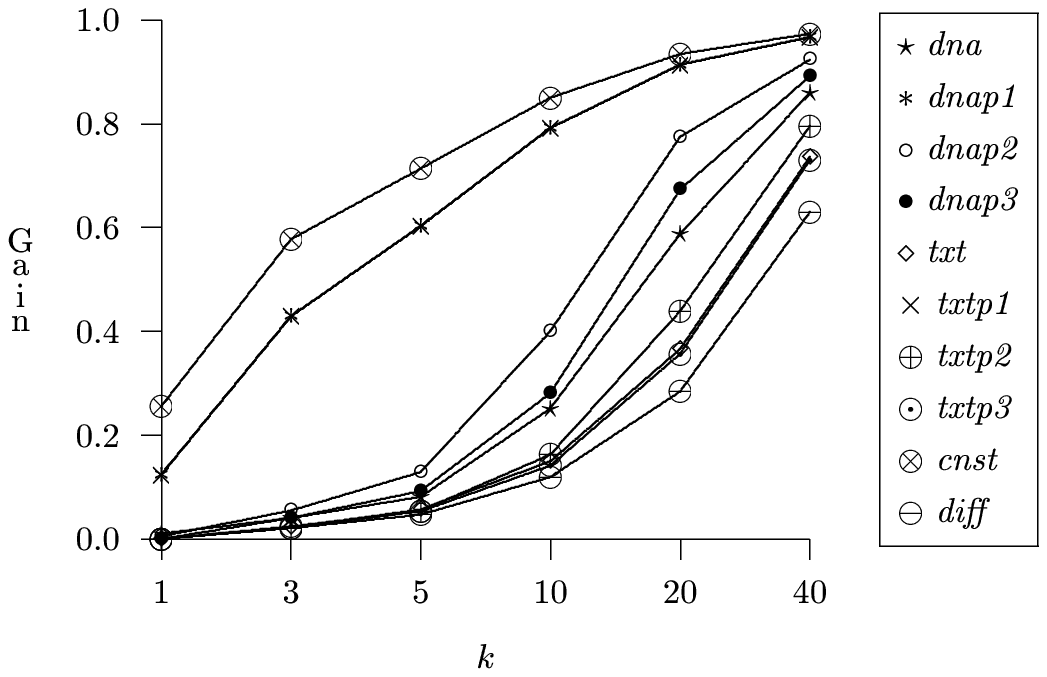}}
\bigskip
\centerline{{\bf Figure 3.} Gains due to the first improvement for $n=50$}
\bigskip\bigskip\bigskip
\endinsert

\topinsert
\centerline{\epsfbox{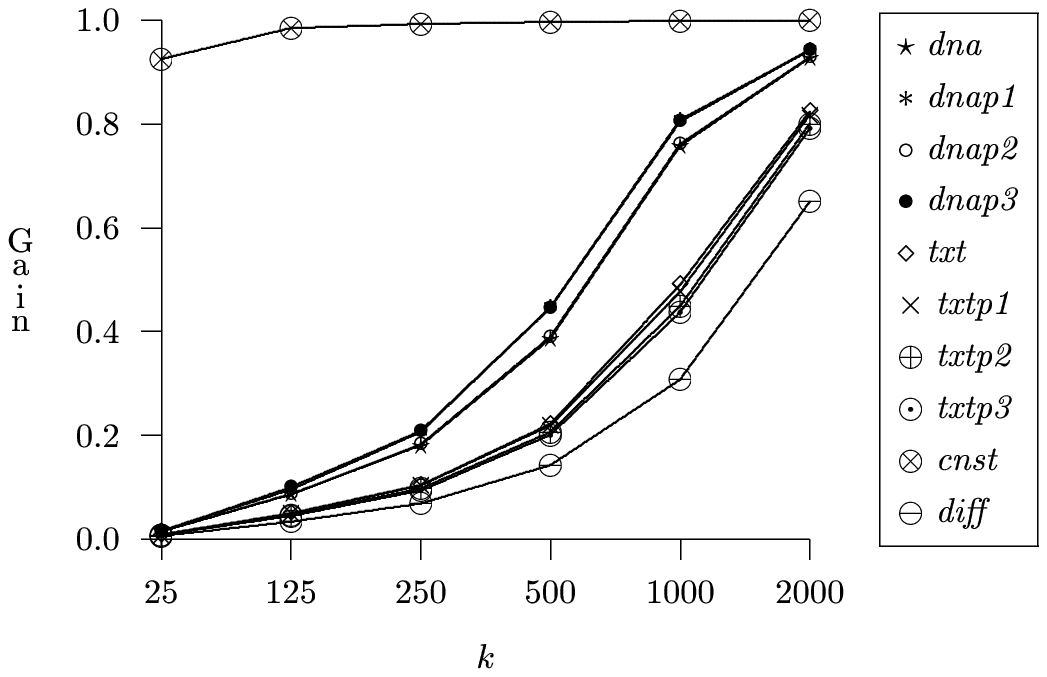}}
\bigskip
\centerline{{\bf Figure 4.} Gains due to the first improvement for $n=2500$}
\bigskip\bigskip\bigskip
\endinsert

\topinsert
\centerline{\epsfbox{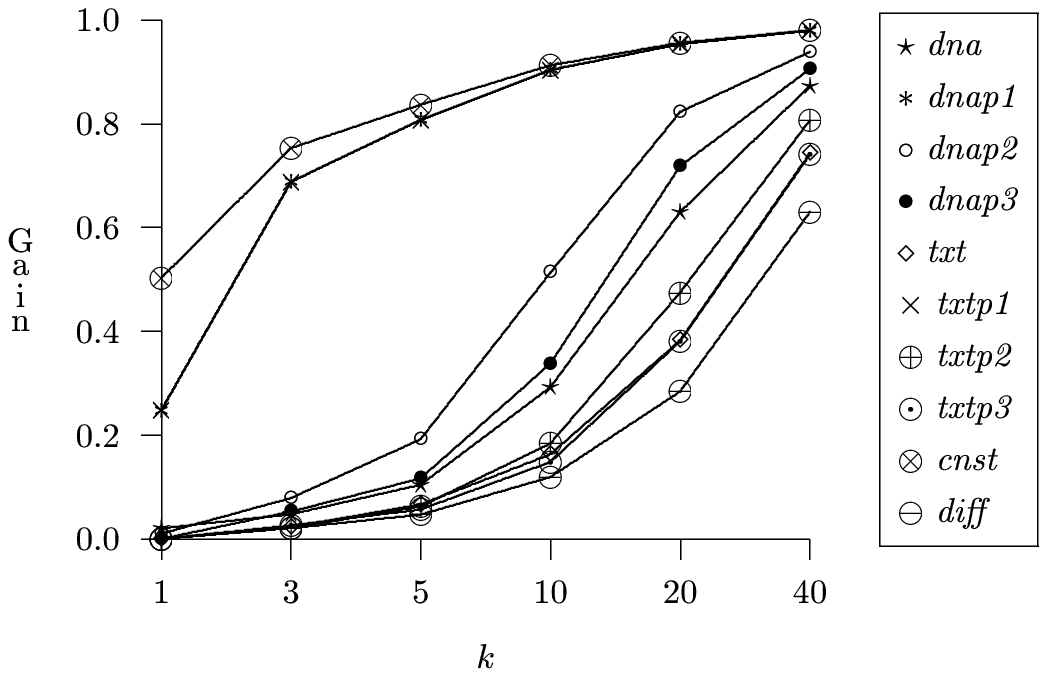}}
\bigskip
\centerline{{\bf Figure 5.} Gains due to the second improvement for $n=50$}
\bigskip\bigskip\bigskip
\endinsert

\topinsert
\centerline{\epsfbox{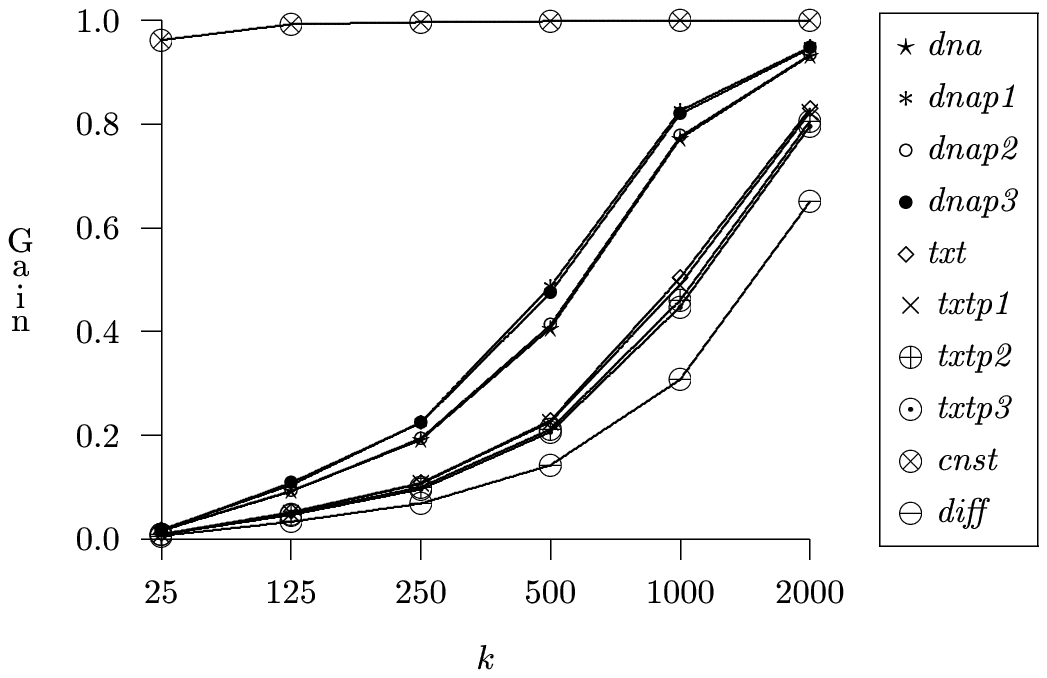}}
\bigskip
\centerline{{\bf Figure 6.} Gains due to the second improvement for $n=2500$}
\bigskip\bigskip\bigskip
\endinsert

\topinsert
\centerline{\epsfbox{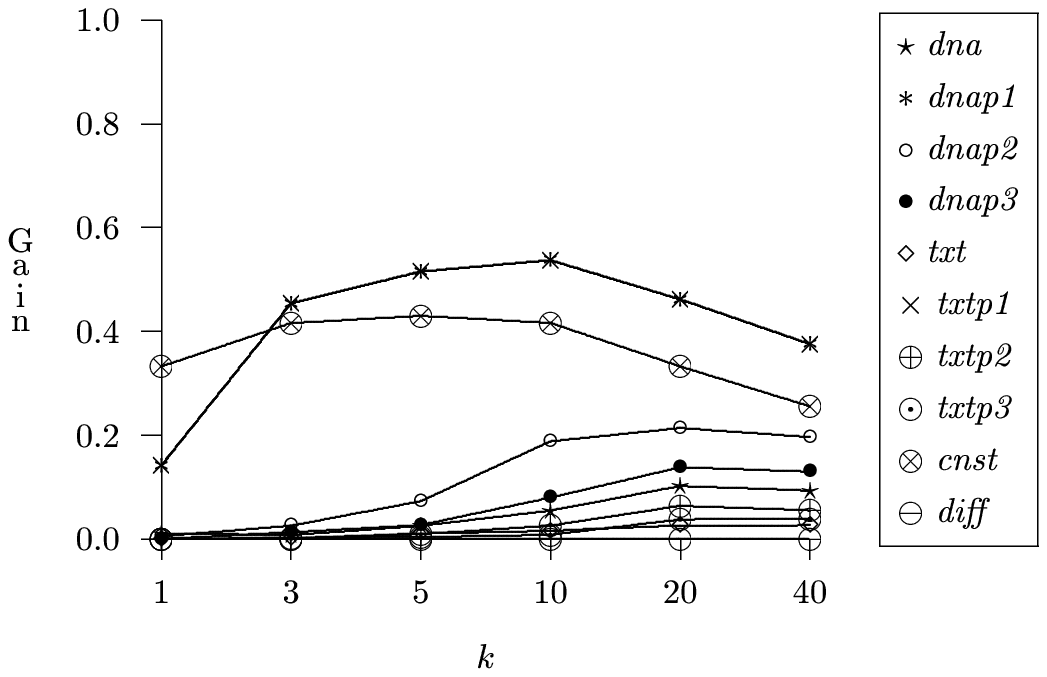}}
\bigskip
\centerline{{\bf Figure 7.} Gains of the second improvement over the first
for $n=50$}
\bigskip\bigskip\bigskip
\endinsert

\topinsert
\centerline{\epsfbox{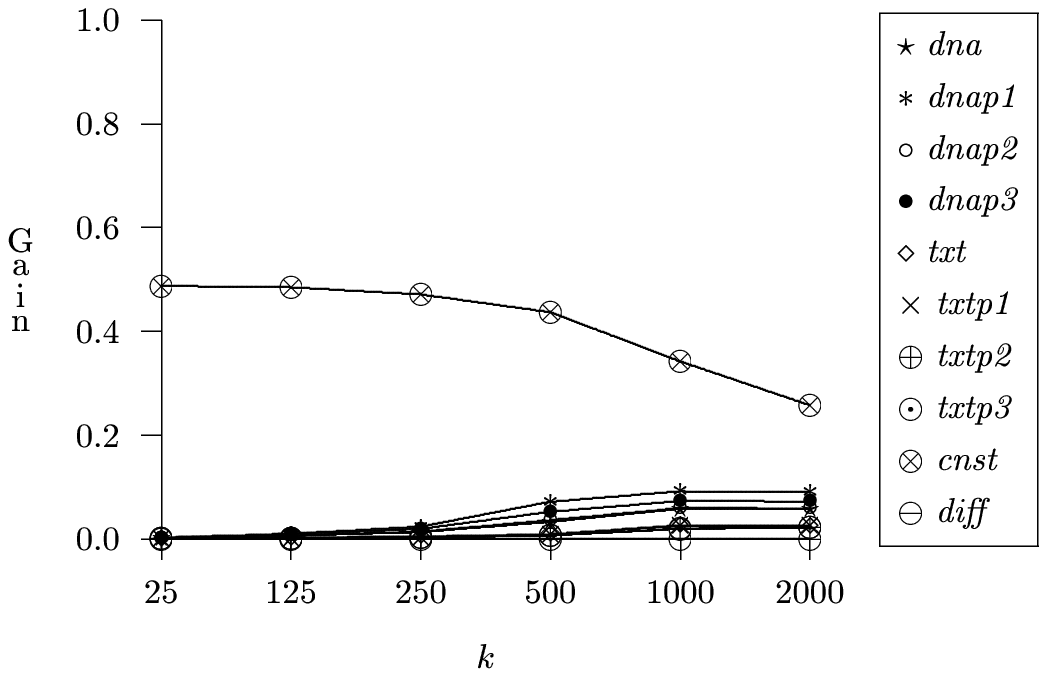}}
\bigskip
\centerline{{\bf Figure 8.} Gains of the second improvement over the first
for $n=2500$}
\bigskip\bigskip\bigskip
\endinsert

\topinsert
\centerline{\epsfbox{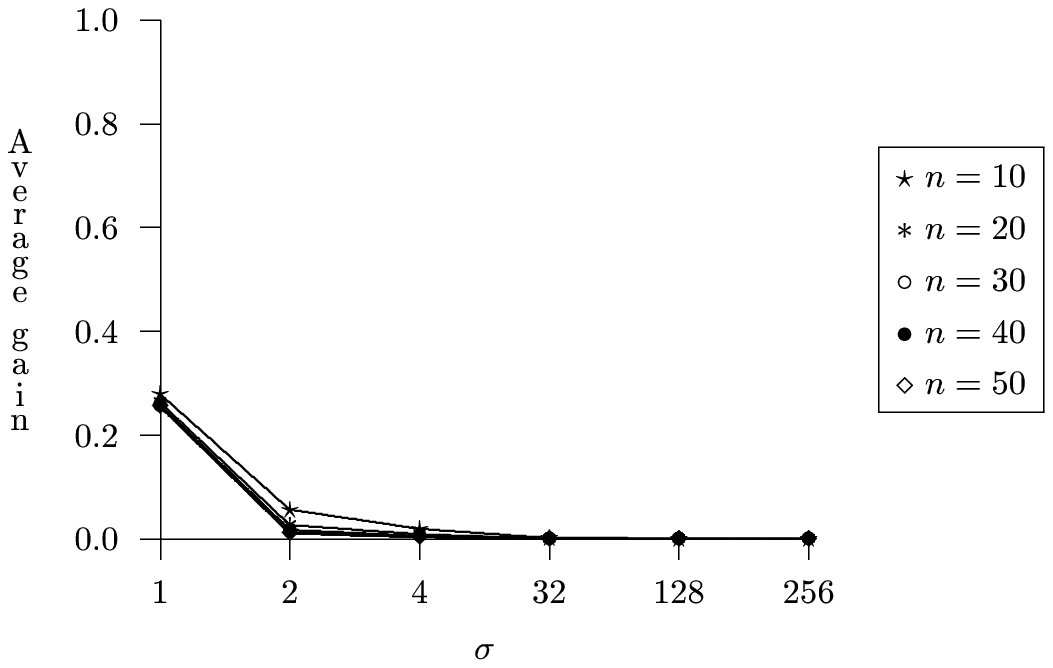}}
\bigskip
\centerline{{\bf Figure 9.} Average gains due to the first improvement for
$k=1$}
\bigskip\bigskip\bigskip
\endinsert

\topinsert
\centerline{\epsfbox{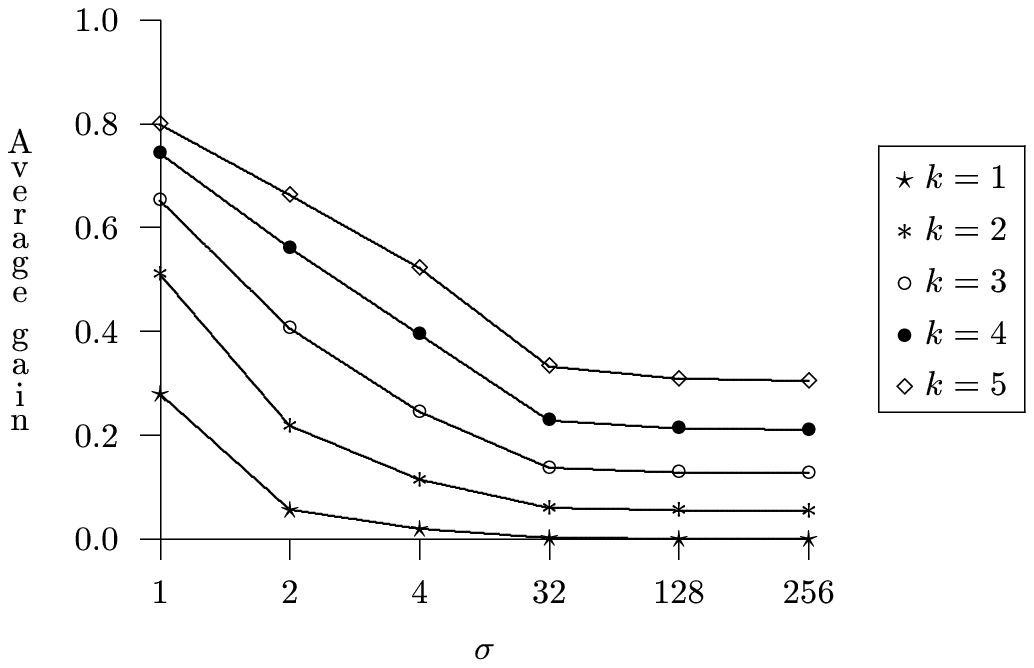}}
\bigskip
\centerline{{\bf Figure 10.} Average gains due to the first improvement for
$n=10$}
\bigskip\bigskip\bigskip
\endinsert

\topinsert
\centerline{\epsfbox{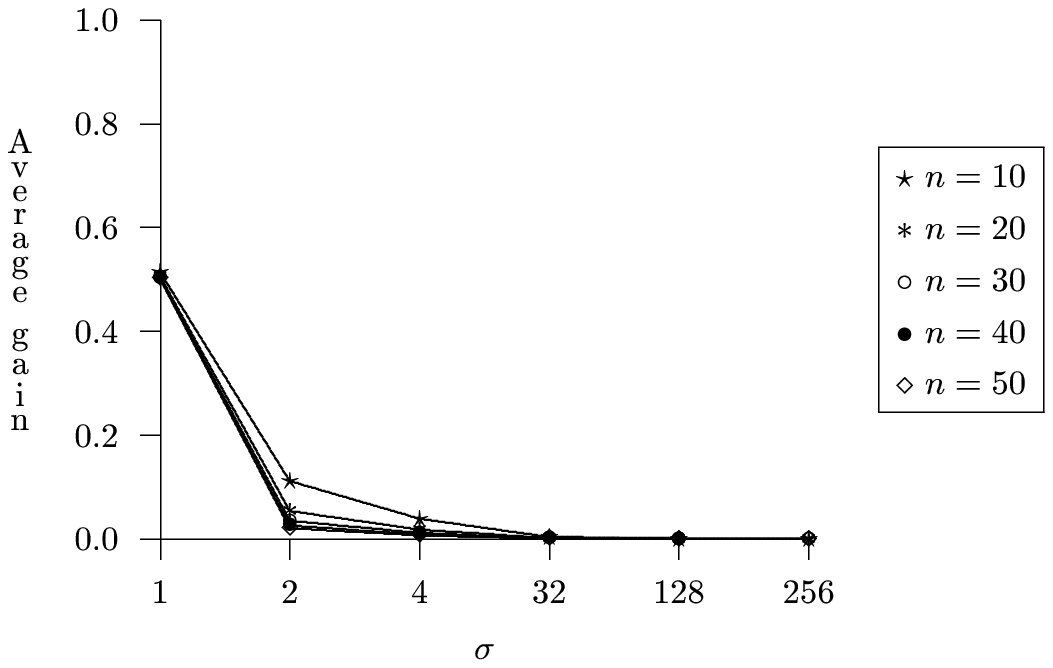}}
\bigskip
\centerline{{\bf Figure 11.} Average gains due to the second improvement for
$k=1$}
\bigskip\bigskip\bigskip
\endinsert

\topinsert
\centerline{\epsfbox{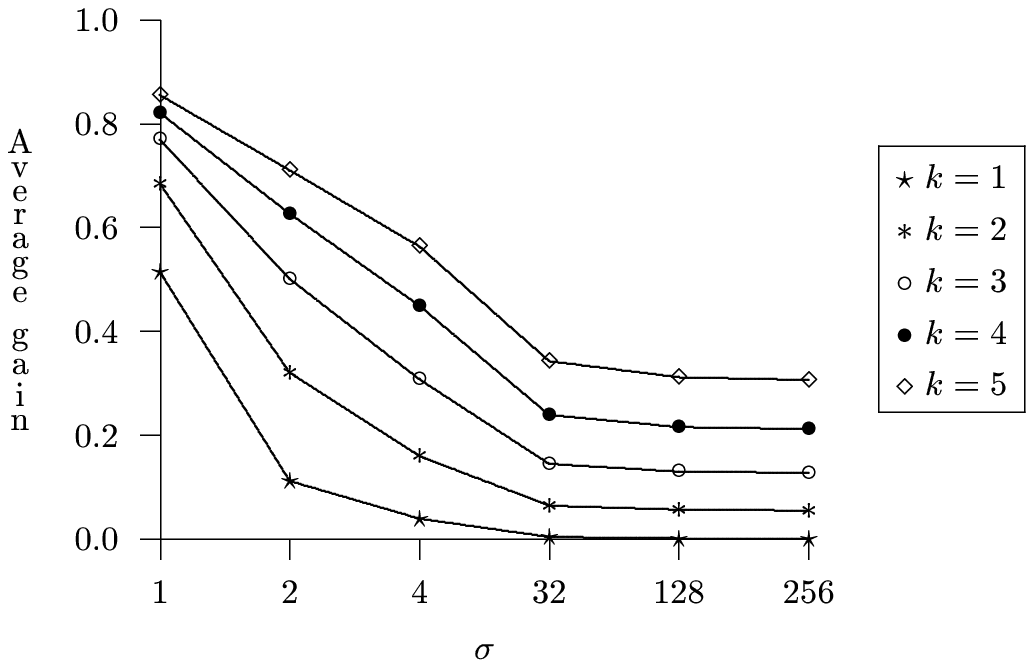}}
\bigskip
\centerline{{\bf Figure 12.} Average gains due to the second improvement for
$n=10$}
\bigskip\bigskip\bigskip
\endinsert

\bye

%% file: 11layout.tex
\magnification=1100
\hsize=6.5 true in
\vsize=9.0 true in
\overfullrule=0pt
\baselineskip=15pt
\tolerance=10000
\hbadness=200

%% file: macro.tex
\font\tensc=cmcsc10
\newfam\scfam
\textfont\scfam=\tensc
\def\sc{\fam\scfam\tensc}
 
\font\bigbf=cmbx10 scaled\magstep1

\font\biggbf=cmbx10 scaled\magstep2

\font\bigggbf=cmbx10 scaled\magstep3

\font\bigit=cmti10 scaled\magstep1

\def\twoldots{\mathinner{\ldotp\ldotp}}

\def\sectionbreak{
\bigskip\vskip\parskip}

\def\bigsectionbreak{
\bigskip\bigskip\vskip\parskip}

\outer\def\beginsection#1\par
{\sectionbreak
\message{#1}\leftline{\bf#1}\nobreak\smallskip\noindent}

\outer\def\bigbeginsection#1\par
{\sectionbreak
\message{#1}\leftline{\bigbf#1}\nobreak\medskip\noindent}

\def\currentsection{\firstmark}

\outer\def\biggbeginsection#1\par
{\bigsectionbreak
\message{#1}\leftline{\biggbf#1}
\mark{#1}\nobreak\bigskip\noindent}

\outer\def\bigitbeginsection#1\par
{\sectionbreak
\message{#1}\leftline{\bigit#1}\nobreak\medskip\noindent}

\outer\def\longbigbeginsection#1 #2\par#3\par
{\sectionbreak
\message{#1 #2 #3}
\halign{##\hfil&##\hfil\cr
{\bigbf#1\ }&{\bigbf#2}\cr
&{\bigbf#3}\cr
}\nobreak\medskip\noindent}

\outer\def\longbiggbeginsection#1 #2\par#3\par
{\bigsectionbreak
\message{#1 #2 #3}
\halign{##\hfil&##\hfil\cr
{\biggbf#1\ }&{\biggbf#2}\cr
&{\biggbf#3}\cr
}\mark{#1 #2 #3}\nobreak\bigskip\noindent}

\outer\def\longbiggbeginappendix#1 #2 #3\par#4\par
{\bigsectionbreak
\message{#1 #2 #3 #4}
\halign{##\hfil&##\hfil\cr
{\biggbf#1\ #2\ }&{\biggbf#3}\cr
&{\biggbf#4}\cr
}\mark{#1 #2 #3 #4}\nobreak\bigskip\noindent}

\def\rightheadline{\hfil{\it\currentsection}\ \ \ \ \ \ {\rm \folio}}

\def\currentchapter{}

\def\leftheadline{{\rm \folio}\ \ \ \ \ \ {\it\currentchapter}\hfil}

\newcount\titlepageno

\def\setheadline{\headline=
{\ifnum\titlepageno=\pageno{\hfil}
\else{\ifodd\pageno{\rightheadline}\else{\leftheadline}\fi}
\fi}}

\def\skipifeven
{\ifodd\pageno{}\else\advancepageno\fi}

\outer\def\beginchapter#1. #2\par
{\vfill\eject\skipifeven\titlepageno=\pageno
\def\currentchapter{Cap\'\i tulo #1. #2}
\topinsert\vskip 0.25\vsize\endinsert
\hrule\medskip
\rightline{\bigggbf Cap\'\i tulo #1}
\medskip
\rightline{\bigggbf#2}
\medskip\hrule\bigskip\bigskip}

\outer\def\shortbeginchapter#1\par
{\vfill\eject\skipifeven\titlepageno=\pageno
\def\currentchapter{#1}
\mark{#1}
\topinsert\vskip 0.25\vsize\endinsert
\hrule\medskip
\rightline{\bigggbf#1}
\medskip\hrule\bigskip\bigskip}

\def\itemitemitem{\par\indent\indent\hangindent3\parindent
\textindent}

\def\LaTeX{{\rm L\kern-.36em\raise.3ex\hbox{\sc a}\kern-.15em%
    T\kern-.1667em\lower.7ex\hbox{E}\kern-.125emX}}

\def\mytilde{\kern-.5pt\lower3pt\hbox{\char'176}\kern.5pt}